\documentclass[prb, twocolumn, floatfix, superscriptaddress, showpacs, longbibliography]{revtex4-2}

\usepackage[utf8]{inputenc}
\usepackage[T1]{fontenc}
\usepackage[main=english]{babel}

\usepackage{mleftright}

\usepackage{amsmath, amsthm, amssymb, mathtools}
\usepackage{bm} 
\usepackage{mathrsfs} 
\usepackage{bbm} 
\usepackage{upgreek} 

\usepackage{xcolor}
\usepackage{graphicx}

\usepackage[colorlinks=true]{hyperref} 
\hypersetup{linkcolor=blue, urlcolor=blue, citecolor=blue, anchorcolor=blue, filecolor=red, menucolor=red}

\newcommand*{\iu}{\mathrm{i}} 
\newcommand*{\Elr}{\mathrm{e}} 

\newcommand*{\abs}[1]{\mleft\lvert {#1} \mright\rvert} 
\newcommand*{\dd}[2][]{\mathop{}\!\mathrm{d}^{#1} {#2}} 
\newcommand*{\var}[2][]{\mathop{}\!\delta_{#1} {#2}} 

\newcommand*{\vdot}{\bm{\cdot}} 
\newcommand*{\vcross}{\bm{\times}} 
\newcommand*{\grad}{\bm{\nabla}} 
\newcommand*{\vb}[1]{\bm{#1}} 
\newcommand*{\rvec}[1]{\overrightarrow{#1}} 
\newcommand*{\lvec}[1]{\overleftarrow{#1}} 
\newcommand*{\vu}[1]{\bm{\hat{#1}}} 

\newcommand*{\LCs}{\upepsilon} 
\newcommand*{\Kd}{\updelta} 
\DeclareMathOperator{\Dd}{\updelta} 

\let\Im\relax
\DeclareMathOperator{\Im}{Im}

\DeclareMathOperator{\Log}{Log}

\begin{document}

\title{Fracton-elasticity duality in twisted moir\'e superlattices}

\author{Jonas Gaa}
\affiliation{Institute for Theory of Condensed Matter, Karlsruhe Institute of Technology, 76131 Karlsruhe, Germany}

\author{Grgur Palle}
\affiliation{Institute for Theory of Condensed Matter, Karlsruhe Institute of Technology, 76131 Karlsruhe, Germany}

\author{Rafael M. Fernandes}
\affiliation{School of Physics and Astronomy, University of Minnesota, Minneapolis, Minnesota 55455, USA}

\author{J\"org Schmalian}
\affiliation{Institute for Theory of Condensed Matter, Karlsruhe Institute of Technology, 76131 Karlsruhe, Germany}
\affiliation{Institute for Quantum Materials and Technologies, Karlsruhe Institute of Technology, 76131 Karlsruhe, Germany}

\begin{abstract}
We  formulate a fracton-elasticity duality
for twisted moir\'e superlattices, taking into account that
they are incommensurate crystals with dissipative phason dynamics. 
From a dual tensor-gauge formulation, as compared to standard crystals,
we identify twice the number of conserved charges that describe
topological lattice defects, namely, disclinations
and a new type of defect that we dub \emph{discompressions}.
The key implication of these conservation laws is that
both glide and climb motions of lattice dislocations are suppressed,
indicating that dislocation networks may become exceptionally stable.
We also generalize our results to other planar incommensurate crystals and quasicrystals.
\end{abstract}

\maketitle

\section{Introduction}
Twisted bilayer graphene (TBG)~\cite{Cao2018,Cao2018b,Balents2020,Andrei2020}
forms an incommensurate moir\'e lattice. The description of its electronic
degrees of freedom, to a very good approximation, can be done using
concepts of periodic crystals, due to the weak scattering
between opposite valleys~\cite{LopesdosSantos2017,Trambly2010,Morell2010,Bistritzer2011}.
However, structurally, there is no periodicity in the system if the
twist angle takes a generic value. As was demonstrated
by Ochoa~\cite{Ochoa2019}, this has profound implications for the elasticity theory of TBG.
Phason modes $\vb{w}\mleft(\vb{x},t\mright)$, which correspond
to acoustic branches of the incommensurate lattice,
dominate lattice vibrations on the scale of the
moir\'e period and give rise to an additional twist stiffness $\kappa$ in the elastic energy
\begin{equation}
E_{\text{el}} \rightarrow E_{\text{el}} + \frac{\kappa}{8} \int \dd[2]{x} \, \mleft[\partial_{x} w_{y} - \partial_{y} w_{x}\mright]^{2}\,. \label{eq:stiffness}
\end{equation}
While in standard elasticity theory such a term is not allowed by rotational
symmetry~\cite{Chaikin2000}, in the case of TBG the adhesive potential between the two
layers gives rise to $\kappa > 0$.
Notice that the elasticity theory of planar quasicrystals~\cite{Levine1985,De1987,Ding1993}
can also be formulated in terms of Eq.~\eqref{eq:stiffness}.
The influence of the twist term and phason
excitations on electron-lattice couplings was discussed
in Refs.~\cite{Ochoa2019,Lian2019}, while their role
in the context of electronic nematicity was analyzed in Ref.~\cite{Fernandes2020}. 

\begin{figure}[t]
\includegraphics[scale=0.4]{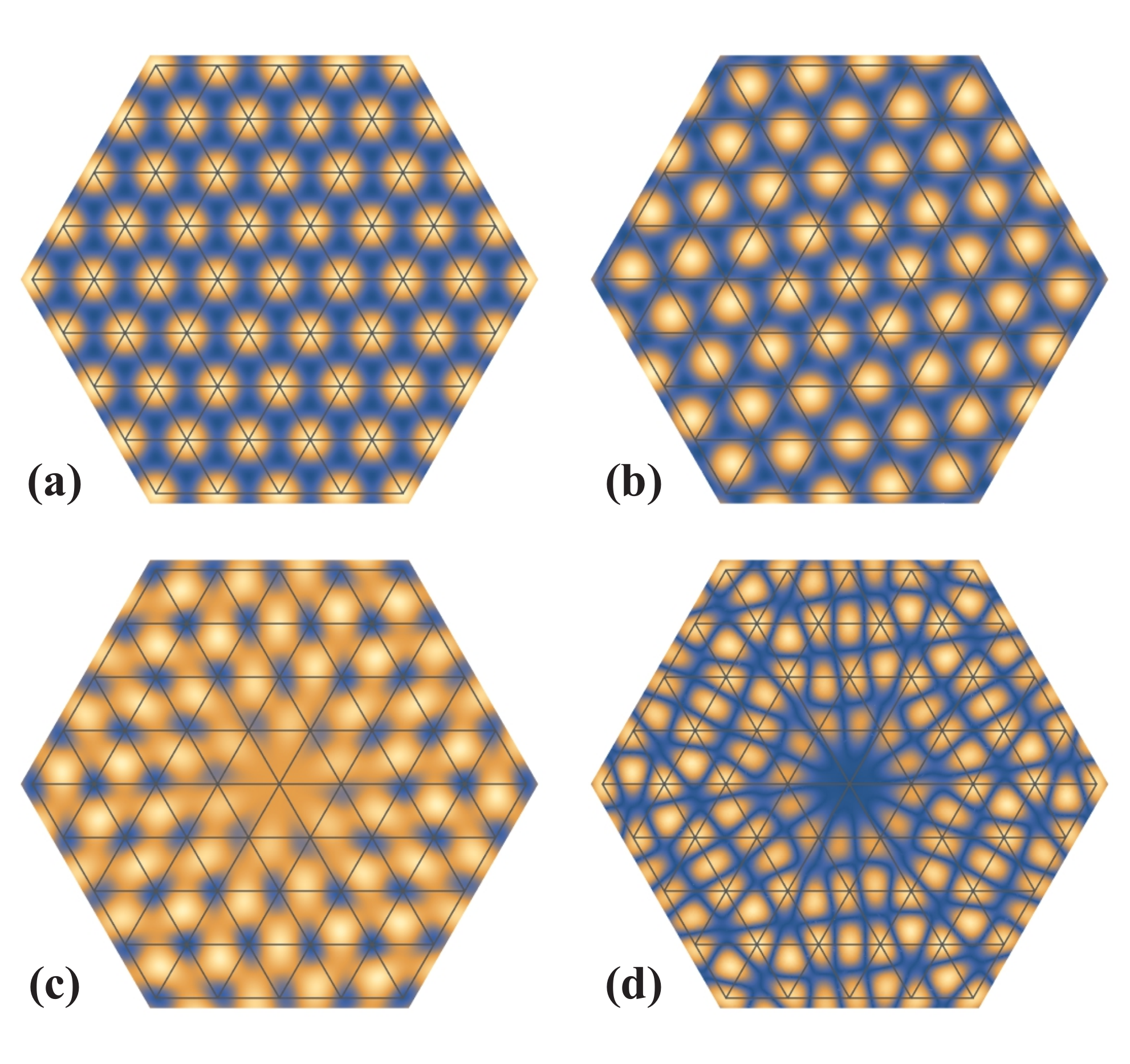}
\caption{Density profile for the triangular moir\'e superlattice, as given by Eq.~\eqref{density},
in the cases without rotation [panel (a), $\varrho_{\theta=0}(\vb{r})$]
and with a rotation of $\theta/(2 \pi) = 0.03$ [panel (b), $\varrho_{\theta}(\vb{r})$].
The phason excitation is illustrated by the difference in density profiles,
$\var{\varrho_{\theta}(\vb{r})} = \varrho_{\theta}(\vb{r}) - \varrho_{\theta=0}(\vb{r})$,
which is plotted in panel (c), and whose absolute value is plotted in panel (d).
The underlying triangular lattice is a guide to the eye.}
\label{fig:phason-density}
\end{figure}

Phasons in TBG have a very straightforward interpretation. 
Ignoring the structural relaxation of the lattice, they
can be identified with the relative displacement between layers~\cite{Ochoa2019}.
To illustrate a twist of the phason mode, we therefore
consider a moir\'e superlattice with sixfold rotational symmetry,
described in terms of the density profile
\begin{equation}
\varrho_{\theta}(\vb{r}) = \sum_{\vb{G}} \abs{\rho_{\vb{G}}} \Elr^{\iu \mleft(\vb{G} \vdot \vb{r} - \phi_{\vb{G}}\mright)}\,. \label{density}
\end{equation}
$\vb{G}$ are reciprocal lattice vectors of the moir\'e superlattice
and $\phi_{\vb{G}} = \vb{w} \vdot \vb{G}$.
In Fig.~\ref{fig:phason-density}, we show the density
profile with $\vb{w} = \theta \vu{z} \vcross \vb{r}$
for $\theta = 0$ and with an excited phason, i.e., for finite rotation of the twist angle $\theta$.
For simplicity, we consider only the leading harmonics, $\vb{G}_1$, $\vb{G}_2$,
and $\vb{G}_3 \equiv - \vb{G}_1 - \vb{G}_2$, with $\abs{\rho_{\vb{G}_i}} \equiv \rho$
and $\vb{G}_{1,2}$ denoting the standard primitive vectors of the triangular lattice.

The dynamics of phasons in incommensurate crystals also differs
from that of standard acoustic phonons by the fact that it is strongly
damped at long times~\cite{Lubensky1985,Baggioli2000}.
This is due to the friction that opposes the
relative motion of the incommensurate mass-density wave.
In usual elasticity, this damping does not occur
because the displacement couples to the generator of translations,
i.e.\ the momentum density. In TBG, however, anharmonic
terms of the adhesion potential between the layers give rise to a finite damping~\cite{Ochoa2021}.

In this paper, we discuss the impact of the twist
stiffness and phason damping on inevitable 
defects of the moir\'e lattice such as dislocations and
disclinations. To this end,
we extend the recently formulated fracton formulation of standard
elasticity~\cite{Pretko2018,Pai2018,Pretko2019} to
incommensurate crystals with overdamped dynamics by developing an appropriate tensor gauge field theory. 
The analogy to a tensor version of electromagnetism allows
us to draw general conclusions about the lattice defect dynamics of TBG. 
The finite twist stiffness of Eq.~\eqref{eq:stiffness}
increases the degrees of freedom compared
to standard elasticity theory, giving rise to
a new type of defect that we call \emph{discompression}.
As a result, defects of the moir\'e lattice,
which can be described in terms of fractons, are governed by an additional conservation law
that dramatically affects the mobility of lattice defects.
In particular, we find that in an incommensurate crystal with phason excitations,
both glide and climb of dislocations are forbidden and
the dynamics of defects is always sub-diffusive. This
implies that dislocation networks 
in twisted bilayer graphene are expected to be exceptionally stable.
We also show that our theory directly applies to other
incommensurate systems such as planar quasicrystals.


The impact of topological lattice defects on the mechanical and electronic properties of graphene and bilayer graphene has been widely discussed~\cite{Louie2010,Kirkland2012,Butz2013}. Lattice disorder has also been recognized as a key ingredient in TBG, with experiments reporting sharp variations of the twist angle and of uniaxial heterostrain over relatively short length scales, which are manifested in local elecronic properties~\cite{Zeldov2020,Bediako2021}. Here, we focus on the topological defects of the emergent moir\'e incommensurate lattice in TBG. While its elastic properties derive from those of the coupled graphene layers, we adopt a coarse-grained low-energy description, as in Ref.~\cite{Ochoa2019}, that does not require mapping the defects of the moir\'e lattice onto individual carbon atoms.

Defects of two-dimensional lattices can be efficiently captured in
terms of a dual description of elasticity theory, where disclinations
emerge as effective charges, dislocations as dipoles of these charges,
and elastic forces between them are transmitted by gauge
fields~\cite{Nelson1979,Young1979,Kleinert1989,Zaanen2004,Cvetkovic2007,Kleinert2008}.
This approach readily demonstrates why dislocations in commensurate lattices can
glide in the direction of the Burgers
vector $\vb{b}$, but not climb perpendicular
to it~\cite{Cvetkovic2007}.
While defects in aperiodic crystals were
discussed in Ref.~\cite{Bohsunng1987,Kleman2003,Kleman2013},
their dynamics remains an open problem.

A very elegant formulation of this duality
was recently achieved in Refs.~\cite{Pretko2018,Pai2018,Pretko2019,Radzihovsky2020a,Gromov2019,Gromov2020b,Nguyen2020}.
It was recognized that the dual formulation
of the usual elasticity theory is, in fact,
a fracton field theory. Fracton phases describe quantum phases of matter with
excitations of restricted mobility~\cite{Nandkishore_review}.
If considered in isolation, a
fracton is immobile, either along certain directions or as a whole.
It can only move collectively, via interactions with other degrees
of freedom. Fractons were initially discussed in the context of non-ergodic
quantum glass models~\cite{Chamon2005} and stabilizer codes for self-correcting
quantum memory~\cite{Haah2011}. More recently, it was recognized that
fractons can be efficiently described in terms of tensor gauge theories~\cite{Pretko2017}.
The restricted mobility of fractons emerges in terms of dipole conservation
laws and gives rise to anomalous, sub-diffusive
hydrodynamics~\cite{Gromov2020,Feldmeier2020,Moudgalya2021,Radzihovsky2020}.

\section{Elasticity theory of twisted bilayer graphene crystals}
We start with the action of elasticity
theory~\cite{Kleinert1989,Chaikin2000},
supplemented by the twist stiffness of Eq.~\eqref{eq:stiffness}.
To include dissipative dynamics, we use
the Schwinger-Keldysh formalism~\cite{Kamenev2011}:
\begin{align}
\begin{aligned}
S &= \frac{1}{2} \int_{\vb{x},t,t' \in \mathcal{C}} w_{i}(\vb{x},t) D^{-1}\mleft(t-t'\mright) w_{i}(\vb{x},t')  \\
&- \frac{1}{2} \int_{\vb{x},t \in \mathcal{C}} \mleft[C_{ij,k\ell} w_{ij}(\vb{x},t) w_{k\ell}(\vb{x},t) + \kappa \vartheta_{w}^2(\vb{x},t)\mright]\,. \label{eq:action}
\end{aligned}
\end{align}
Here, $\int_{\vb{x},t \in \mathcal{C}} = \int_{\mathcal{C}} \dd{t} \int_{V} \dd[2]{x}$
indicates that the time integration has to be performed
on the round-trip Keldysh contour, while the spatial
integration goes over the crystal volume $V$.
$w_{ij} \equiv \frac{1}{2} \mleft(\partial_i w_{j} + \partial_j w_{i}\mright)$
is the symmetric phason strain, the elastic constants $C_{ij,k\ell}$ are
summarized in Ref.~\cite{Ochoa2019}, and the bond angle variation
is $\vartheta_{w} \equiv \frac{1}{2} \LCs_{ij} \partial_i w_{j}$.
Damping enters through the nonlocal-in-time contribution
\begin{equation}
 D^{-1}\mleft(t-t'\mright) = - \Dd\mleft(t-t'\mright) \partial_{t'}^{2} + \gamma\mleft(t-t'\mright)\,,  
\end{equation}
where we assume Ohmic damping $\Im \gamma^{R}(\omega) = \gamma_{0} \omega$
for the retarded self energy. Without damping, the equation of motion
become elastic wave equations, while the dynamics
becomes diffusive at large $\gamma_0$. For further details on the Keldysh formlism, see Appendix~\ref{app:A}.

We shall use a notation that will prove efficient in 2D.
In it, we always start with lower index vectors or tensors $A_{i}$.
The raising of their indices we define as a contraction
with the Levi-Civita symbol:
\begin{equation}
A^{i} \equiv \LCs_{ij} A_{j}\,.
\end{equation}
From $A_{i} = \mleft(A_{x}, A_{y}\mright)$ and $A^{i} \equiv \LCs_{ij} A_{j}$ 
we see that the upper index vector $A^{i} = \mleft(A_{y}, - A_{x}\mright)$
is the lower index vector rotated in the clockwise direction by $\uppi/2$. This
ensures that $\vb{A} \vdot \vb{B} = A_{i} B_{i} = A^{i} B^{i}$.
On the other hand,
$\mleft(\vb{A} \vcross \vb{B}\mright)_{z} = A_{i} B^{i} = - A^{i} B_{i}$
and in particular $A_i A^i = 0$. The divergence and two-dimensional curl
may likewise be expressed in this notation as
\begin{align}
\grad \vdot \vb{A} &= \partial_{i} A_{i}\,, \nonumber \\
\mleft(\grad \vcross \vb{A}\mright)_{z} &= \partial_{i} A^{i} = - \partial^{i} A_{i}\,.
\end{align}
For single-valued functions $f$, $\partial_{i} \partial^{i} f = 0$. For
multivalued functions, the partial derivative do not commute at the branch cuts.
The most important benefit of this notation, reminiscent of the
van der Waerden spinor notation~\cite{vdWaerden1928},
is that the correspondence between elasticity and tensor
electromagnetism~\cite{Pretko2018,Pai2018,Pretko2019,Radzihovsky2020a}
amounts to simply raising all the indices.

\subsection{Dual gauge theory}

Building on the approach of Refs.~\cite{Pretko2018,Pai2018,Pretko2019},
the first step in formulating a dual description is to
introduce the fields
$\partial_t w_{i} \to \pi_i$, $C_{ij,k\ell} w_{k\ell} \to \sigma_{ij}$,
and $\kappa \vartheta_{w} \to M$
through a series of Hubbard-Stratonovich transformations.
They can be associated with momentum, stress, and torque, respectively.
This yields the real-time action
\begin{align}
S &= \frac{1}{2} \int_{\vb{x}, t \in \mathcal{C}} \mleft[C_{ij,k\ell}^{-1} \sigma_{ij}(\vb{x},t) \sigma_{k\ell}(\vb{x},t) + \kappa^{-1} M^{2}(\vb{x},t)\mright] \nonumber \\
&\quad - \frac{1}{2} \int_{\vb{x},t,t' \in \mathcal{C}} \pi_{i}(\vb{x},t) \Gamma\mleft(t-t'\mright)\pi_{i}(\vb{x},t') + S_{w}\,, \label{afterHStransf}
\end{align}
where the retarded  version of $\Gamma$ is
$\Gamma^{R}(\omega) = \omega^{2} / \mleft(\omega^{2} + \iu \gamma_{0} \omega\mright)$.  
The last term
\begin{equation}
S_{w} = \int_{\vb{x},t \in \mathcal{C}} \mleft[\pi_{i} \mleft(\partial_{t} w_{i}\mright) - \sigma_{ij} w_{ij} - M \vartheta_{w}\mright] \label{eq:stress-strain-term}
\end{equation}
still contains the original phason field.

The next step
is to decompose $w_{i} = \tilde{w}_{i} + w_{i}^{(s)}$
into a smooth part $\tilde{w}_{i}$ 
and a singular part $w_{i}^{(s)}$ due to topological defects.
After integrating out the smooth part, 
we obtain the constraint
\begin{equation}
\partial_t \pi_j = \partial_i \Sigma_{ij}\,, \label{constraint}
\end{equation}
which resembles Newton's second law with a non-symmetric stress
\begin{equation}
\Sigma_{ij}=\sigma_{ij} + \frac{1}{2} \LCs_{ij} M\,,
\end{equation}
consisting of
the symmetric stress tensor $\sigma_{ij}$ and the torque $M$.
Whereas $\sigma_{ij}$ also appears in ordinary crystals,
the torque only arises due
to the finite twist stiffness.

The last step is to introduce (tensorial) gauge potentials that enforce
the constraint~\eqref{constraint}:
\begin{align}
\begin{aligned}
\Sigma_{ij} &= - \partial_t A_{ij} - \partial^i \phi_j\,, \\
\pi_j &= - \partial_i A_{ij}\,.
\end{aligned}
\end{align}
These gauge potentials are invariant under the gauge transformations
\begin{align}
\begin{aligned}
A_{ij} &\mapsto A_{ij} + \partial^{i} \Lambda_{j}\,, \\
\phi_{j} &\mapsto \phi_{j} - \partial_t \Lambda_{j}\,.
\end{aligned} \label{gauge_transformation}
\end{align}
The stress-strain coupling Eq.~\eqref{eq:stress-strain-term} now takes
the familiar form of a charge/current-potential coupling:
\begin{equation}
S_{w} = \int_{\vb{x},t \in \mathcal{C}} \mleft(A_{ij} J_{ij} - \phi_{j} Q_{j}\mright)\,. \label{lagrangian_gauge}
\end{equation}
Due to the tensorial nature of the gauge fields,
the charge density 
\begin{equation}
Q_{j} = \partial^{i} \partial_{i} w_{j}^{(s)}
\end{equation} 
is a vector
and the corresponding current density,
\begin{equation}
J_{ij} = \mleft(\partial_i \partial_t - \partial_t \partial_i\mright) w_{j}^{(s)}\,,
\end{equation}
a tensor.
Demanding the invariance of~\eqref{lagrangian_gauge} under
the gauge transformations
yields the continuity equation:
\begin{equation}
\partial_{t} Q_{j} + \partial^{i} J_{ij} = 0\,. \label{continuity_p}
\end{equation}

\subsection{Electromagnetic analogy}
Let us develop, in analogy to Ref.~\cite{Pretko2018,Pai2018,Pretko2019,Radzihovsky2020a} for standard elasticity,
a physical illuminating electromagnetic analogy of
the conserved vector charge $Q_j$. Consider a single
charge 
\begin{equation}
Q_{j}(\vb{x},t) = q_{j} \Dd\mleft(\vb{x}-\vb{r}(t)\mright)
\end{equation}
at
position $\vb{r}(t)$ that moves with velocity $\vb{v}(t) = \dot{\vb{r}}(t)$.
Eq.~\eqref{continuity_p} is then fulfilled by
\begin{equation}
J_{ij}(\vb{x},t) = v^{i} q_{j} \Dd\mleft(\vb{x}-\vb{r}(t)\mright)\,.    
\end{equation}
Inserting this into the action~\eqref{lagrangian_gauge}
yields the Lorentz force
\begin{equation}
F_{i} = \var{S_{w}} / \var{r_{i}} = - \mleft(\Sigma^{i}_{~j} + v^{i} \pi_{j}\mright) q_{j}\,.
\end{equation}
The problem behaves like the electrodynamics
of non-symmetric tensor electric fields $ \Sigma^{i}_{~j}$
and vector magnetic fields $\pi_{i}$. 
This  analogy also helps us
understand why the overdamped dynamics of the phasons does not spoil
the entire gauge description. The dissipative propagator $\Gamma(t-t')$ for
$\pi_i(t)$ in Eq.~\eqref{afterHStransf} translates,
in standard electromagnetism, to a frequency-dependent
magnetic permeability $\Gamma(\omega)$, which clearly
does not violate the gauge description of electromagnetism. 

Besides the effects of dissipation, a crucial
new aspect of the incommensurate moir\'e lattice elasticity
is the vector continuity equation~\eqref{continuity_p}
that can be rephrased in terms of \emph{two} conserved scalar densities:
\begin{align}
\begin{aligned}
\rho^{(\ell)} &= \partial_{j} Q_{j}\,, \\
\rho^{(t)} &= \partial^{j} Q_{j}\,,
\end{aligned}
\end{align}
with continuity equations
\begin{align}
\begin{aligned}
\partial_{t} \rho^{(\ell)} + \partial_{j} \partial^{i} J_{ij} &= 0\,, \\
\partial_{t} \rho^{(t)} + \partial^{j} \partial^{i} J_{ij} &= 0\,.
\end{aligned} \label{continuity_div_p}
\end{align}
The transverse density $\rho^{(t)}$ also appears
in standard elasticity, where it was identified
as the density of disclinations, with $Q_i$ denoting
the Burgers vector density~\cite{Zaanen2004,Pretko2018}.
The longitudinal density $\rho^{(\ell)}$
is the new conserved charge present only in incommensurate crystals.

\subsection{Interpretation of the charges}
The charges of the elastic gauge theory
are topological lattice defects that make
the displacement field $w_{i}$ multivalued~\cite{Kleinert2008}.
We can write the singular part of a dislocation displacement field as
\begin{equation}
\vb{w}^{(s,Q)}(\vb{x}) = \frac{\Im \Log(z)}{2 \uppi} \, \vb{b}\,, \label{us_disloc}
\end{equation}
where $z = x_1 + \iu x_2$, $\Log(z)$ is the principal branch of the logarithm
with a branch cut along the negative $x_1$ axis,
and $\vb{b} = b_1 \vu{e}_1 + b_2 \vu{e}_2$ is the Burgers vector.
At the origin the partial derivatives do not commute, yielding
$Q_j(\vb{x}) = b_j \Dd(\vb{x})$, which confirms that the
vector charge is the Burgers vector density.

Next, we construct a disclination from a Dirac string of dislocations.
By integrating $\int_{-\infty}^{0} \dd{x_1'} \, \vb{w}^{(s,Q)}(x_1-x_1', x_2)$,
ignoring all except the singular
parts, and setting $b_1=0$ and
$b_2 \neq 0$, we obtain
\begin{equation}
\vb{w}^{(s,t)}(\vb{x}) = - b_2 \frac{\Im\mleft[z \Log(z)\mright]}{2 \uppi} \, \vu{e}_2\,, \label{us_discli}
\end{equation}
with $Q_j = b_2 \Kd_{2j} \Theta(-x_1) \Dd(x_2)$,
$\rho^{(\ell)} = b_2 \Theta(-x_1) \Dd'(x_2)$, and
$\rho^{(t)} = b_2 \Dd(\vb{x})$.
This confirms that $\rho^{(t)}$ is the density of disclinations.
In addition, the result for $\rho^{(l)}$ implies that a disclination
also corresponds to a Dirac string along the negative $x_1$-axis
made of dipoles ($\sim \Dd'(x_2)$) of the longitudinal charge.

\begin{figure}[t]
\centering
\includegraphics[scale=0.8]{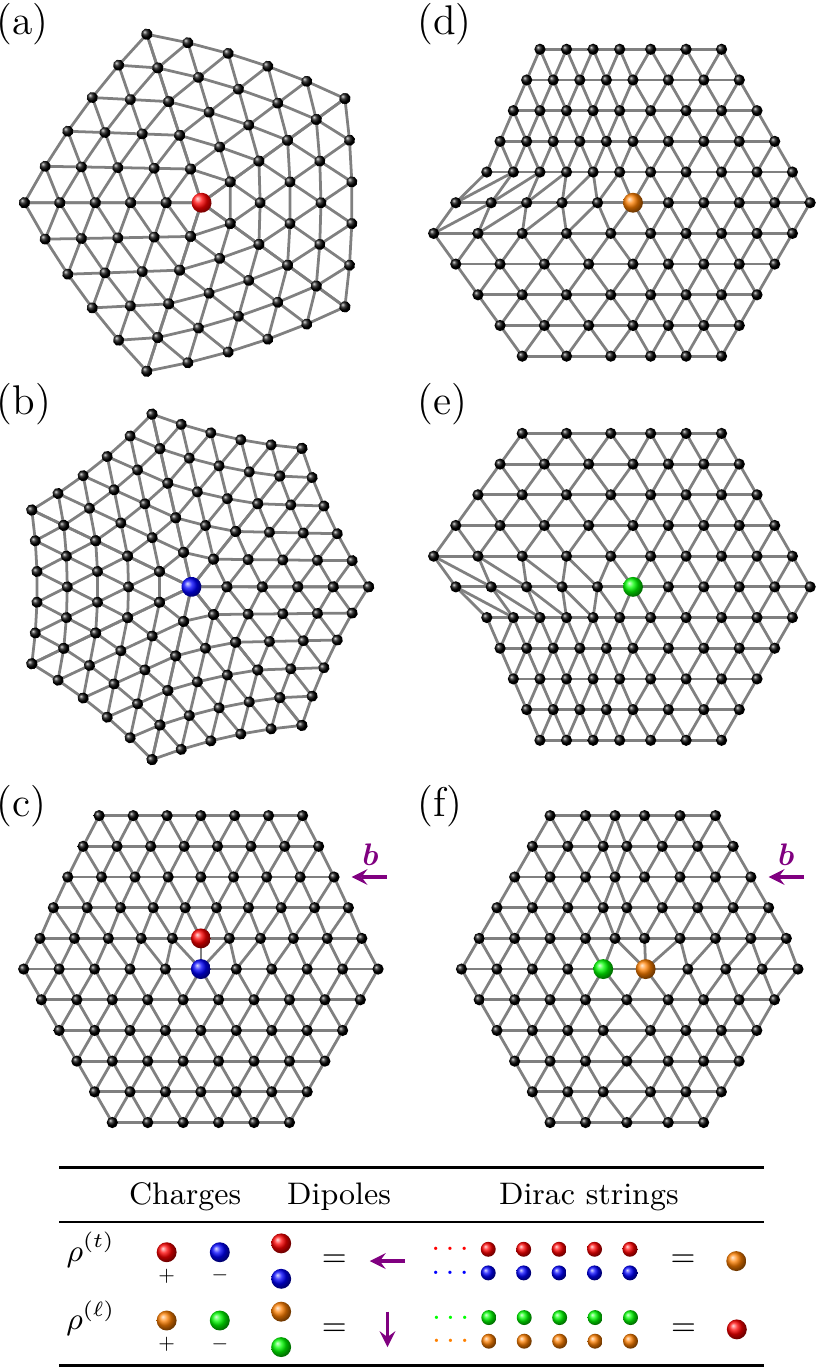}
\caption{Various defects for a triangular lattice and the relations between them.
(a) Positive disclination ($b_2 = 2 / \sqrt{3}$).
(b) Negative disclination ($b_2 = - 2 / \sqrt{3}$).
(c) Dislocation as a dipole of disclinations.
(d) Positive discompression ($b_1 = 1/2$).
(e) Negative discompression ($b_1 = -1/2$).
(f) Dislocation as a dipole of discompressions.
The table summarizes the fusion rules that explain how to represent charges in terms of Dirac strings of dipoles of other charges.}
\label{fig:defects}
\end{figure}

Similarly, we may consider a Dirac string of dislocations
whose Burgers vectors point along the branch cut,
instead of perpendicular to it 
($b_1 \neq 0$ and $b_2 = 0$),
\begin{equation}
\vb{w}^{(s,\ell)}(\vb{x}) = - b_1 \frac{\Im\mleft[z \Log(z)\mright]}{2 \uppi} \, \vu{e}_1\,. \label{us_discompress}
\end{equation}
The charges are $Q_j = b_1 \Kd_{1j} \Theta(-x_1) \Dd(x_2)$,
$\rho^{(\ell)} = - b_1 \Dd(\vb{x})$, and
$\rho^{(t)} = b_1 \Theta(-x_1) \Dd'(x_2)$.
Since $\vb{w}^{(s,\ell)}$ points parallel to the Dirac string,
and its jump in value is proportional to the distance from
the origin, it represents an abrupt change in the strain.
We shall thus call this defect a \emph{discompression}.
Below Eq.~\eqref{us_discli} we showed that a disclination
is equivalent to a Dirac string of discompression dipoles.
The result for $\rho^{(t)}$ implies, in turn, that a discompression
equals a Dirac string of disclination dipoles. These ``fusion rules'' for the
various defects are sketched in the lower part of Fig.~\ref{fig:defects}.

In Fig.~\ref{fig:defects}, we show these defects for a triangular lattice.
Disclinations and discompressions are energetically
very expensive as both come along with macroscopic
regions of compression or bond-angle mismatch.
More formally, both are forbidden as single
charges by an appropriate generalization of Weingarten's theorem~\cite{Kleinert2008}.
Hence, a moir\'e lattice is charge neutral
with regards to $\rho^{(t,\ell)}$, while their
dipoles and higher moments remain perfectly valid excitations. 
Fig.~\ref{fig:defects}(c) illustrates the known fact that a dislocation
corresponds to a dipole of disclinations (panels (a) and (b))
with dipole moment perpendicular to the Burgers vector $\vb{b}$.
Our analysis shows that a dislocation [Fig.~\ref{fig:defects}(f)] can
also emerge from two discompressions of opposite charge (panels (d) and (e)).
The dipole moment is now oriented along the Burgers vector.
Single discompressions of Fig.~\ref{fig:defects}(d) and (e) can be
generated in terms of a Volterra process
where we cut along the negative $x_1$-axis and displace matter 
above and below the cut in opposite directions parallel to the cut,
before glueing back together. This is distinct from disclinations
of Fig.~\ref{fig:defects}(a) and (b) where the displacement
is perpendicular to the cut and opens a wedge.

\subsection{Consequences of the continuity equations}
Eq.~\eqref{continuity_p} implies the
conservation of the total Burgers vector.
Let us analyze this vector conservation law in terms of
the continuity equations~\eqref{continuity_div_p} of the scalar densities
that both contain two spatial derivatives. This implies, in addition to the conservation
of the total disclination and discompression charges,
that both of their total dipole moments are conserved as well:
\begin{equation}
\frac{\dd{}}{\dd{t}} \int \dd[2]{x} \  x_k \, \rho^{(\ell,t)} = 0\,. \label{conservation_disclin_di}
\end{equation}
Both charges are therefore immobile fractons.
Only correlated movements at fixed dipole density are allowed.

\subsection{Energetics and equations of motion}
Thus far we analyzed conservation laws that follow from
the gauge invariance of the dual theory. To understand the
actual dynamics of  defects requires, however,  an analysis of
the energetics of vector charges. To this end we determine
the equations of motion of the gauge fields that
mediate the stresses. The Euler-Lagrange equation of $A_{ij}$ is:
\begin{equation}
\partial_{t} \tilde{\Sigma}_{ij}(t) = \int \dd{t'} \, \Gamma^{R}\mleft(t-t'\mright) \partial_{i}\pi_{j}(t') - J_{ij}(t)\,, \label{eq:A_ij-EL}
\end{equation}
with $\tilde{\Sigma}_{ij} = C_{ij,k\ell}^{-1} \sigma_{k\ell} + \LCs_{ij} \kappa^{-1} M$.
It is the analog of Ampere's law in electrodynamics.
The analog of Gauss's law $\partial^{i} \tilde{\Sigma}_{ij} = Q_{j}$
follows from the variation of the potential $\phi_j$ and demonstrates
that $Q_j$ is indeed the source of stress.

We can use Ampere's law
to determine the dynamics of the quadrupolar moments.
To achieve this, we first contract the
Euler-Lagrange equation~\eqref{eq:A_ij-EL} with $\Kd_{ij}$
and $\LCs_{ij}$. This gives
\begin{align}
\partial_t C_{ii,k\ell}^{-1} \sigma_{k\ell} &= \partial_i (\Gamma^{R} \pi_i) - J_{ii}\,, \label{contr_delta} \\
\frac{2}{\kappa} \partial_t M &= \partial_i (\Gamma^{R} \pi^i) - J^{~i}_i\,, \label{contr_epsilon}
\end{align}
where
\begin{equation}
\mleft(\Gamma^{R} \pi_j\mright)(\vb{x},t) = \int \dd{t'} \, \Gamma^{R}\mleft(t - t'\mright) \pi_j(\vb{x},t')\,.
\end{equation}

Now we use the conservation laws~\eqref{continuity_div_p},
partially integrate the quadrupolar moments twice, and then
use the equations of motion. This yields
\begin{align}
\frac{\dd{}}{\dd{t}} \int_V \abs{\vb{x}}^2 \rho^{(t)} ~ &\overset{\mathclap{\eqref{continuity_div_p}}}{=} ~ - \int_V x_k x_k \partial^{j} \partial^{i} J_{ij} \label{sup:quad-der1} \\
&= - 2 \int_V J_{ii} + \int_V \partial_{j} \mleft(2 x_i J^{ji} - \abs{\vb{x}}^2 \partial_{i} J^{ij}\mright) \notag \\
&\overset{\mathclap{\eqref{contr_delta}}}{=} ~ 2 \int_V \partial_t C_{ii,k\ell}^{-1} \sigma_{k\ell} \notag \\
&+ \int_V \partial_{j} \mleft(2 x_i J^{ji} - \abs{\vb{x}}^2 \partial_{i} J^{ij} - 2 \Gamma^R \pi_{j}\mright) \notag
\end{align}
and
\begin{align}
\frac{\dd{}}{\dd{t}} \int_V \abs{\vb{x}}^2 \rho^{(\ell)} \label{sup:quad-der2} ~ &\overset{\mathclap{\eqref{continuity_div_p}}}{=} ~ - \int_V x_k x_k \partial_{j} \partial^{i} J_{ij} \\
&= - 2 \int_V J^{~i}_{i} + \int_V \partial_{j} \mleft(\abs{\vb{x}}^2 \partial_{i} J^{i}_{~\,j} - 2 x_i J^{j}_{~\,i}\mright) \notag \\
&\overset{\mathclap{\eqref{contr_epsilon}}}{=} ~ \frac{4}{\kappa} \int_V \partial_t M \notag \\
&+ \int_V \partial_{j} \mleft(\abs{\vb{x}}^2 \partial_{i} J^{i}_{~\,j} - 2 x_i J^{j}_{~\,i} - 2 \Gamma^R \pi^{j}\mright)\,. \notag
\end{align}
Finally, we rewrite the sources by exploiting the
Euler-Lagrange equations of the action~\eqref{afterHStransf},
$C_{ij,k\ell}^{-1} \sigma_{k\ell} = w_{ij}$ and $M = \kappa \vartheta_{w}$,
and thus obtain
\begin{equation}
\frac{\dd{}}{\dd{t}} \int \dd[2]{x} \, \abs{\vb{x}}^2 \rho^{(\ell,t)} = \frac{\dd{}}{\dd{t}} S^{(\ell,t)}\,. \label{conservation_quad}
\end{equation}
The source terms for disclination quadrupoles is
\begin{equation}
    S^{(t)} = 2 \int \dd[2]{x} \, \partial_i w_{i}
\end{equation}
and corresponds to a costly volume change of the system. In turn, 
\begin{equation}
    S^{(\ell)} = 4 \int \dd[2]{x} \, \vartheta_{w}
\end{equation}
for discompression
quadrupoles corresponds to a global twist.
These quadrupoles can easily be
interpreted if one considers an isolated dislocation $Q_j(\vb{x}) = b_j \Dd\mleft(\vb{x} - \vb{r}(t)\mright)$ whose $\int \dd[2]{x} \, \abs{\vb{x}}^2 \rho^{(t)} = 2 b^j r_j(t)$
and $\int \dd[2]{x} \, \abs{\vb{x}}^2 \rho^{(\ell)} = - 2 b_j r_j(t)$.
Thus the quadrupolar moments are the components of the dislocation position
parallel and perpendicular to the Burgers vector. 
Eq.~\eqref{conservation_quad} was obtained up to boundary terms. The boundary terms vanish or, more physically, cannot be
changed by local operations that act in the bulk of the medium.
Thus~\eqref{conservation_quad} expresses how
in the bulk the sources of changing
quadrupolar moments are the energetically costly
compression and disorientation of the crystal.

In our analysis we freely used expressions
that are valid only on-shell, the various conservation laws (\ref{continuity_p}, \ref{continuity_div_p}, \ref{conservation_disclin_di}, \ref{conservation_quad}),
equations of motion (\ref{eq:eomdisplfull}, \ref{eq:A_ij-EL}, $C_{ij,k\ell}^{-1} \sigma_{k\ell} = \partial_i w_{j}$, $M = \kappa \vartheta_{w}$),
expressions for the sources $S^{(t)}$ and  $S^{(\ell)}$,
etc., can all be formulated as rigorous statements
that apply to the respective path-integral averages.
This is proved by using the Schwinger-Dyson equations.

From Eq.~\eqref{conservation_quad} we may therefore conclude that a climb, i.e.\ the
motion perpendicular to $b_j$ of a dislocation,
is accompanied by an energetically costly change in the volume of the crystal.
In addition, a glide, i.e.\ the motion parallel to $b_j$ of a dislocation, 
is accompanied by an energetically costly change of the orientation of the crystal. 
While the first result holds in any crystal,
the second is due to the non-zero twist stiffness
of incommensurate ones. In its presence the motion of dislocations is forbidden
entirely instead of just being restricted to one direction.

\subsection{Sub-diffusive hydrodynamics}

Thus far we have analyzed the motion of isolated dipoles and quadrupoles. The situation
becomes more complicated for higher-order multipoles. However, using the hydrodynamic
description of fractons~\cite{Gromov2020,Feldmeier2020,Moudgalya2021,Radzihovsky2020} one readily finds that there
cannot be a Fick's law where gradients of the vector
charges yield currents (i.e.\ $J_{ij} \propto \partial_i Q_j$ or $\propto \partial^i Q_j$),
leading to ordinary diffusion of all charges. If this were true, the mere existence
of scalar charges would induce a current $J_{ii} \sim \rho^{(\ell,t)}$, which is not allowed.
Expanding in gradients, the leading 
symmetry-allowed term connects $J_{ij}$ with
a third spatial derivative of $Q_{j}$.
This gives rise to a coupled dynamics of the two
scalar densities that form two sub-diffusive
modes with dispersions $\omega = \iu B_{\pm} k^4$.


\section{Generalization to generic planar quasicrystals and incommensurate crystals}
\label{sec:III}
Finally, we demonstrate that our results are not restricted to TBG and
apply directly to planar quasicrystals and incommensurate crystals with two length scales.
To this end  we relate the elastic gauge theory with twisting term
considered in the main text
to the elasticity theory of two-dimensional quasicrystals.
In these systems, the low-energy elastic variables are
doubled compared to periodic crystals~\cite{Levine1985,De1987,Ding1993}.
In addition to the phonons $\vb{u}$, there are phasons $\vb{w}$ that can
be shown to have a twist stiffness like in Eq.~\eqref{eq:stiffness}.
The defects of the system are characterized
by two Burgers vectors $\vb{b}_{u} = \oint \dd{\vb{u}}$ and $\vb{b}_{w} = \oint \dd{\vb{w}}$~\cite{De1987}.
Our results imply that pure displacement dislocations can glide,
whereas those that involve $\vb{b}_{w}$ cannot.
This remains true even in the presence of phonon-phason coupling.

A symmetry analysis of the problem yields the following result for
the elastic energy~\cite{Levine1985,De1987,Ding1993}:
\begin{equation}
\mathcal{E}_{\text{el}} = \frac{1}{2} C_{ij,k\ell} u_{ij} u_{k\ell} + R_{ij,k\ell} u_{ij} w_{k\ell} + \frac{1}{2} K_{ij,k\ell} w_{ij} w_{k\ell}\,,
\end{equation}
where $u_{ij} = \frac{1}{2} \mleft(\partial_{i} u_{j} + \partial_{j} u_{i}\mright)$
is the usual symmetric strain tensor describing phonons and $w_{ij} = \partial_{i} w_{j}$
is the non-symmetric phason tensor.

For the moment, let us focus on the case of a twelvefold symmetric planar quasicrystals
that has the special property of no phonon-phason mixing, i.e., $R_{ij,k\ell} = 0$~\cite{Ding1993}.
Because of the high rotational symmetry, the elastic constants $C_{ij,k\ell}$
have the same form as those of isotropic media:
\begin{equation}
C_{ij,k\ell} = \lambda \, \Kd_{ij} \Kd_{k\ell} + \mu \, \mleft(\Kd_{ik} \Kd_{j\ell} + \Kd_{i\ell} \Kd_{jk}\mright)\,. \label{sup:eq:C_ijkl}
\end{equation}
On the other hand,
\begin{equation}
K_{ij,k\ell} = K_{1} \Kd_{ik} \Kd_{j\ell} + K_{2} \mleft(\Kd_{ij} \Kd_{k\ell} - \Kd_{i\ell} \Kd_{jk}\mright) + \tilde{K}_{3} \LCs^{S}_{ij} \LCs^{S}_{k\ell}\,, \label{sup:eq:K_ijkl}
\end{equation}
where $\LCs^{S}_{11} = \LCs^{S}_{22} = 0$ and $\LCs^{S}_{12} = \LCs^{S}_{21} = 1$.
We note that $K_{ij,k\ell}$ satisfies only the major index symmetry
$K_{ij,k\ell} = K_{k\ell,ij}$, but not the minor ones
$K_{ij,k\ell} \neq K_{ji,k\ell}$ and $K_{ij,k\ell} \neq K_{ij,\ell k}$.

To make contact with our system~\eqref{eq:action},
we merely have to recognize that $K_{ij,k\ell}$
can be rewritten as
\begin{equation}
K_{ij,k\ell} = \tilde{C}_{ij,k\ell} + \frac{\kappa}{4} \LCs_{ij} \LCs_{k\ell}\,, \label{sup:decomposition}
\end{equation}
where
\begin{equation}
\tilde{C}_{ij,k\ell} = \tilde{\lambda} \, \Kd_{ij} \Kd_{k\ell} + \tilde{\mu} \, \mleft(\Kd_{ik} \Kd_{j\ell} + \Kd_{i\ell} \Kd_{jk}\mright) + \tilde{K}_{3} \LCs^{S}_{ij} \LCs^{S}_{k\ell}
\end{equation}
is the minor-index-symmetric part that
satisfies $\tilde{C}_{ij,k\ell} = \tilde{C}_{ji,k\ell} = \tilde{C}_{ij,\ell k}$
and has $\tilde{\lambda} = K_{2}$ and $\tilde{\mu} = \frac{1}{2} \mleft(K_{1} - K_{2}\mright)$.
The minor-index-antisymmetric part plays the role of a twisting term with stiffness
\begin{equation}
\kappa = 2 \mleft(K_{1} + K_{2}\mright)\,. \label{sup:eff-kappa}
\end{equation}
Thus the phonon displacement field $u_{i}$ behaves like in the elasticity theory
of a conventional crystal, while the phason $w_{i}$ is governed by an elastic energy identical
to the one of twisted bilayer graphene, given in Eq.~\eqref{eq:stiffness} or~\eqref{eq:action}.
Hence all our conclusions about the mobility of defects of the phason modes carry over.

The situation is more complicated for quasicrystals with fivefold or eightfold symmetry.
In these cases, $C_{ij,k\ell}$ is still isotropic, as in~\eqref{sup:eq:C_ijkl}, and
$K_{ij,k\ell}$ still has the form~\eqref{sup:eq:K_ijkl},
although with $\tilde{K}_3 = 0$ in the case of fivefold symmetry,
and can therefore be decomposed
according to~\eqref{sup:decomposition}. Most significantly,
a phonon-phason coupling of the form
\begin{equation}
R_{ij,k\ell} = R \mleft(\Kd_{i1} - \Kd_{i2}\mright) \mleft[\Kd_{ij} \Kd_{k\ell} + \Kd_{ik} \Kd_{j\ell} - \Kd_{i\ell} \Kd_{jk}\mright]
\end{equation}
is allowed. Consequently, one has to redo the whole gauge formulation.

The end results, however, turn out to be insensitive to phonon-phason coupling.
Specifically, for the phonon field one finds
the usual results~\cite{Pretko2018,Pai2018,Pretko2019}:
\begin{align}
\partial_{t} \rho^{(t)}_{u} + \partial^{j} \partial^{i} J_{uij}^{S} &= 0\,, \\
\frac{\dd{}}{\dd{t}} \int \dd[2]{x} \  x_k \, \rho^{(t)}_{u} &= 0\,, \\
\frac{\dd{}}{\dd{t}} \int \dd[2]{x} \, \abs{\vb{x}}^2 \rho^{(t)}_{u} &= \frac{\dd{}}{\dd{t}} S^{(t)}_{u}\,,
\end{align}
where $Q_{uj} = \partial^{i} \partial_{i} u_{j}^{(s)}$,
$J_{uij} = \mleft(\partial_i \partial_t - \partial_t \partial_i\mright) u_{j}^{(s)}$,
$\rho^{(t)}_{u} = \partial^{j} Q_{uj}$,
$J_{uij}^{S} = \frac{1}{2} \mleft(J_{uij} + J_{uji}\mright)$, and
$S^{(t)}_{u} = 2 \int \dd[2]{x} \, \partial_i u_i$. For the
phason fields one similarly finds that \eqref{continuity_p},
\eqref{continuity_div_p}, \eqref{conservation_disclin_di},
and \eqref{conservation_quad} continue to hold, but with the
appropriate $Q_{wj} = \partial^{i} \partial_{i} w_{j}^{(s)}$,
$J_{wij} = \mleft(\partial_i \partial_t - \partial_t \partial_i\mright) w_j^{(s)}$,
$\rho^{(\ell)}_{w} = \partial_{j} Q_{wj}$, $\rho^{(t)}_{w} = \partial^{j} Q_{wj}$,
$S^{(t)}_{w} = 2 \int \dd[2]{x} \, \partial_i w_i$, and
$S^{(\ell)}_{w} = 4 \int \dd[2]{x} \, \vartheta_w$.
Therefore, the glide of defects that have a finite
phason Burgers vectors $\vb{b}_{w} = \oint \dd{\vb{w}}$
is suppressed, and the parameter that controls the supression
is the effective twisting stiffness~\eqref{sup:eff-kappa}.

\emph{Details of the phonon-phason coupled gauge theory:}
It is convenient to introduce an additional index $\mu,\nu \in \{u,w\}$
that differentiates the phonon and phason displacement fields. That way,
$u_{ui} = u_{i}$, $u_{wi} = w_{i}$, $u_{uij} = u_{ij}$, $u_{wij} = w_{ij}$, and
\begin{gather}
C_{uij,uk\ell} = C_{ij,k\ell}\,, \quad C_{wij,wk\ell} = K_{ij,k\ell}\,, \\
C_{uij,wk\ell} = C_{wk\ell,uij} = R_{ij,k\ell}\,.
\end{gather}
The appropriate generalization of the action~\eqref{eq:action} we may now write as
\begin{align}
\begin{aligned}
S &= \frac{1}{2} \int_{\vb{x},t,t' \in \mathcal{C}} u_{\mu i}(t) D^{-1}_{\mu}\mleft(t-t'\mright) u_{\mu i}(t') \\
&\qquad - \frac{1}{2} \int_{\vb{x},t \in \mathcal{C}} C_{\mu ij,\nu k\ell} u_{\mu ij} u_{\nu k\ell}\,.
\end{aligned}
\end{align}
Next, we introduce the Hubbard-Stratonovich fields
$\partial_t u_{\mu i} \to \pi_{\mu i}$ and $C_{\mu ij,\nu k\ell} u_{\nu k\ell} \to \Sigma_{\mu ij}$,
integrate out the regular parts of $u_{\mu i}$, and then enforce the
constraints $\partial_t \pi_{\mu j} = \partial_i \Sigma_{\mu ij}$
through gauge potentials:
\begin{align}
\begin{aligned}
\Sigma_{uij} &= - \partial_t A_{uij}^{S} - \partial^{i} \partial^{j} \phi_{u}\,, \\
\pi_{uj} &= - \partial_{i} A_{uij}^{S}\,,
\end{aligned} \\
\begin{aligned}
\Sigma_{wij} &= - \partial_t A_{wij} - \partial^{i} \phi_{wj}\,, \\
\pi_{wj} &= - \partial_{i} A_{wij}\,,
\end{aligned}
\end{align}
where $A_{uij}^{S} = A_{uji}^{S}$ is symmetric. This yields the action
\begin{align}
\begin{aligned}
S &= - \frac{1}{2} \int_{\vb{x},t,t' \in \mathcal{C}} \pi_{\mu i}(t) \Gamma_{\mu}\mleft(t-t'\mright)\pi_{\mu i}(t') \\
&\qquad + \frac{1}{2} \int_{\vb{x}, t \in \mathcal{C}} C_{\mu ij,\nu k\ell}^{-1} \Sigma_{\mu ij} \Sigma_{\nu k\ell} + S_{uw}\,,
\end{aligned} \label{sup:afterHStransf-2}
\end{align}
where $\Gamma_{\mu}\mleft(t-t'\mright) = \lvec{\partial_{t}} D_{\mu}\mleft(t-t'\mright) \rvec{\partial_{t'}}$ and
\begin{align}
\begin{aligned}
S_{uw} = \int_{\vb{x},t \in \mathcal{C}} \big[&A_{uij}^{S} J_{uij}^{S} + \phi_{u} \rho^{(t)}_{u} \\
&+ A_{vij} J_{vij} - \phi_{vj} Q_{vj}\big]\,,
\end{aligned}
\label{sup:charge-potential-coupling}
\end{align}
with the previously defined charges and current densities.

In this action we see why the phonon-phason coupling does
not affect the argument leading to the quadrupolar
conservation laws. On the one hand, the charge conservation
laws are a consequence of the local gauge symmetry and
from the form of the source term~\eqref{sup:charge-potential-coupling}
we immediately see that they are unaffected by $C_{\mu ij,\nu k\ell}^{-1}$
or $R_{ij,k\ell}$. On the other hand, let us
take a look at the Euler-Lagrange
equations of $A_{\mu ij}$ that allowed for an additional
partial integration in~\eqref{sup:quad-der1} and~\eqref{sup:quad-der2}:
\begin{equation}
\partial_{t} \tilde{\Sigma}_{\mu ij}(t) = \int \dd{t'} \, \Gamma^{R}_{\mu}\mleft(t-t'\mright) \partial_{i}\pi_{\mu j}(t') - J_{\mu ij}(t)\,.
\end{equation}
Once one expresses $\tilde{\Sigma}_{\mu ij} = C_{\mu ij,\nu k\ell}^{-1} \Sigma_{\nu k\ell}$
in terms of the original fields $u_{\mu ij}$,
one finds that $\tilde{\Sigma}_{\mu ij} = u_{\mu ij}$
according to the Euler-Lagrange equations of $\Sigma_{\mu ij}$.
Thus $C_{\mu ij,\nu k\ell}$ again drop out of the argument.
Therefore, $C_{\mu ij,\nu k\ell}$ formally influence the argument
only through the symmetry properties that they
entail for $\Sigma_{\mu ij}$. Physically, however,
$C_{\mu ij,\nu k\ell}$ also set the energy
scales of glide and climb suppressions.

\section{Conclusions}
The dual formulation of elasticity theory that
exploits the concept of fractons allows for rather general insights
into the mobility of dissipative incommensurate crystals, such as twisted bilayer graphene.
In particular, we find that these systems have, in distinction to usual crystals, completely
immobile dislocations in the low defect density limit
that become sub-diffusive in the high-density limit. 
Hence, while the electronic properties of graphene can, to a very good approximation, be treated using concepts of periodic crystals~\cite{LopesdosSantos2017,Trambly2010,Morell2010,Bistritzer2011}, the incommensurate nature of the material is much more visible in its mechanical properties. 
In order to estimate whether the effects discussed here are quantitatively relevant,  we follow~\cite{Ochoa2019}. Here, $\kappa$ is shown to be of the order of $10 \text{eV}/\text{nm}^2  \sin\mleft(\Theta/2\mright)$. For twist angles $\Theta \sim 1^{\circ}$ this yields $\kappa \sim 0.1 \text{eV}/\text{nm}^2$. Hence, on the length scale of the moir\'e crystal $\sim 10 \text{nm}$, this stiffness is clearly relevant in a wide temperature regime.

Our results are not unique to twisted bilayer graphene. As we demonstrate in section~\ref{sec:III}, they can be generalized to generic planar quasicrystals and incommensurate crystals. The lack of mobility of defects stabilizes networks of defects, such as the soliton network discussed in Ref.~\cite{Ochoa2019}. Another implication is that these incommensurate systems should mechanically be rather brittle. 

The power of the formalism that led to the results of this work
is drawn primarily from the deep intuition that
follows from the analogy to electromagnetism,
 including multipole expansions or the electromagnetism of dissipative media.
An interesting open question is how the properties of these defects,
and particularly the discompressions, impact the global mechanical
properties of TBG, as well as its local electronic spectrum, where stable defect configurations on the electronic spectrum are expected to lead to localized bound state formation. For isolated disclinations and discompressions one might even expect topologically protected bound states in the electronic spectrum~\cite{Peterson2021}.

\emph{Note added:} After this work was completed, we became aware of interesting related works concerned with a fracton description~\cite{Surowka2021} and topologically protected defect motion~\cite{Else2021} of quasicrystals.

\begin{acknowledgments}
We thank H. Ochoa for fruitful discussions. J.~G.\ was supported by the Deutsche Forschungsgemeinschaft (DFG, German Research Foundation) project SCHM 1031/12-1.
G.~P.\ and J.~S.\ were supported by the DFG - TRR 288 - 422213477 Elasto-Q-Mat (project A07).
R.M.F. was supported by the U.S. Department of Energy, Office of Science, Basic Energy Sciences, under award no. DE-SC0020045.
R.M.F. also acknowledges a Mercator Fellowship (TRR 288 - 422213477) from the German Research Foundation (DFG).
\end{acknowledgments}

\appendix

\section{Schwinger-Keldysh formulation of the elasticity theory}
\label{app:A}

In order to include damping of fractons, we perform the analysis on
the Keldysh contour~\cite{Kamenev2011}. We consider the generating functional:
\begin{equation}
W\mleft[h_{i}\mright] = \int \mathcal{D}w \, \Elr^{\iu S\mleft[w\mright] - \iu \int_{\mathcal{C}} h_{i}\mleft(t\mright) w_{i}\mleft(t\mright)}\,.
\end{equation}
Here, $\int_{x,t\in\mathcal{C}} \cdots = \int_{\mathcal{C}} \dd{t} \int_{V} \dd[2]{x} \cdots$
indicates that the time integration has to be performed on the round-trip
Keldysh contour, while the spatial integration goes over the volume $V$.
The action of the problem is given in Eq.~\eqref{eq:action}, where  
the damping term enters through the nonlocal-in-time contribution
\begin{equation}
D^{-1}\mleft(t-t'\mright) = - \Dd\mleft(t-t'\mright) \partial_{t'}^{2} + \gamma\mleft(t-t'\mright)\,,
\end{equation}
with friction self energy term $\gamma\mleft(t\mright)$. Let $w_{i}^{+}\mleft(t\mright)$
and $w_{i}^{-}\mleft(t\mright)$ refer to the phason modes on the upper and
lower contour, respectively. Transforming the Keldysh degrees of freedom
to quantum and classical fields 
\begin{equation}
w_{i}^{c,q}\mleft(t\mright) = \frac{1}{\sqrt{2}}\mleft(w_{i}^{+}\mleft(t\mright) \pm w_{i}^{-}\mleft(t\mright)\mright)\,,
\end{equation}
the self energy as function of frequency takes the form 
\begin{equation}
\gamma\mleft(\omega\mright) = \begin{pmatrix}
0 & \gamma^{R}\mleft(\omega\mright) \\
\gamma^{A}\mleft(\omega\mright) & \gamma^{K}\mleft(\omega\mright)
\end{pmatrix}\,.
\end{equation}
We assume Ohmic damping
\begin{equation}
\Im \gamma^{R}\mleft(\omega\mright) = \gamma_{0} \omega \Elr^{- \abs{\omega} / \omega_{c}}
\end{equation}
with upper cutoff $\omega_{c}$. The real part of $\gamma^{R}\mleft(\omega\mright)$
is determined via a Kramers-Kronig transformation, where constant terms
$\gamma^{R}\mleft(0\mright)$ have to be subtracted. In addition,
\begin{equation}
\gamma^{K}\mleft(\omega\mright) = - 2 \coth\mleft(\frac{\omega}{2T}\mright) \Im \gamma^{R}\mleft(\omega\mright)
\end{equation}
follows from the fluctuation-dissipation theorem. Hence, the retarded
Fourier transform is at low energies given by
\begin{equation}
D^{R}\mleft(\omega\mright) = \frac{1}{\omega^{2} + \iu \gamma_{0} \omega}\,.
\end{equation}
At low frequencies, the damping term is the dominant one.
The equations of motion for the displacement fields $w_{i}$ are
\begin{align}
\begin{aligned}
\frac{\partial^{2} w_{j}}{\partial t^2} &= \int \dd{t'} \, \gamma^{R}\mleft(t-t'\mright) w_{j}(t') \\
&\qquad + \partial_{i} \mleft[C_{ij,k\ell} w_{k\ell} + \tfrac{1}{2} \LCs_{ij} \kappa \vartheta_{w}\mright]\,.
\end{aligned} \label{eq:eomdisplfull}
\end{align}
For strong damping and long times 
$\int \dd{t'} \, \gamma^{R}\mleft(t-t'\mright) \to - \gamma_0 \partial_t$,
and the dynamics is diffusive.
When we perform the Hubbard-Stratonovich transformations,
we obtain Eq.~\eqref{afterHStransf} with
\begin{equation}
\Gamma\mleft(t-t'\mright) = \overleftarrow{\partial}_{t} D\mleft(t-t'\mright) \overrightarrow{\partial}_{t'}\,.
\end{equation}
The retarded version of $\Gamma$ becomes 
\begin{equation}
\Gamma^{R}\mleft(\omega\mright) = \frac{\omega^{2}}{\omega^{2} + \iu \gamma_{0}\omega}\,.
\end{equation}
In the limit $\gamma_{0} \to 0$, $\Gamma^{R}\mleft(\omega\mright) \to 1$.


\begin{thebibliography}{100}
\bibitem{Cao2018}Y. Cao, V. Fatemi, A. Demir, S. Fang, S. L. Tomarken,
J. Y. Luo, J. D. Sanchez-Yamagishi, K. Watanabe, T. Taniguchi, E.
Kaxiras, R. C. Ashoori, and P. Jarillo-Herrero, \emph{Correlated insulator
behaviour at half-filling in magic-angle graphene superlattices},
Nature \textbf{556}, 80 (2018).

\bibitem{Cao2018b}Y. Cao, V. Fatemi, S. Fang, K. Watanabe, T. Taniguchi,
E. Kaxiras, and P. Jarillo-Herrero, \emph{Unconventional superconductivity
in magic-angle graphene superlattices}, Nature \textbf{556}, 43 (2018).


\bibitem{Balents2020} L. Balents, C. R. Dean, D. K. Efetov, and A. Young, \emph{Superconductivity and strong correlations in moir\'e flat bands}. Nature Phys. \textbf{16}, 725 (2020).

\bibitem{Andrei2020}E. Y. Andrei and A. H. MacDonald, \emph{Graphene
bilayers with a twist}, Nat. Mater. \textbf{19}, 1265 (2020).

\bibitem{LopesdosSantos2017}J. M. B. Lopes dos Santos, N. M. R. Peres,
and A. H. Castro Neto, \emph{Graphene bilayer with a twist: Electronic
structure}, Phys. Rev. Lett. \textbf{99}, 256802 (2007).

\bibitem{Trambly2010}G. Trambly de Laissardi\`ere, D. Mayou, and L.
Magaud, \emph{Localization of Dirac electrons in rotated graphene
bilayers}, Nano Lett. \textbf{10}, 804 (2010).

\bibitem{Morell2010}E. Su\'arez Morell, J. D. Correa, P. Vargas, M.
Pacheco, and Z. Barticevic, \emph{Flat bands in slightly twisted bilayer
graphene: Tight-binding calculations}, Phys. Rev. B \textbf{82}, 121407
(2010).

\bibitem{Bistritzer2011}R. Bistritzer and A. H. MacDonald, \emph{Moir\'e
bands in twisted double-layer graphene}, Proc. Natl. Acad. Sci. U.S.A.
\textbf{108}, 12233 (2011).

\bibitem{Ochoa2019}H. Ochoa, \emph{Moir\'e-pattern fluctuations and
electron-phason coupling in twisted bilayer graphene,} Phys. Rev.
B \textbf{100}, 155426 (2019).

\bibitem{Chaikin2000}P. M. Chaikin and T. C. Lubensky, \emph{Principles
of Condensed Matter Physics} (Cambridge University Press, Cambridge,
2000).

\bibitem{Levine1985}D. Levine, T. C. Lubensky, S. Ostlund, S. Ramaswamy,
P. J. Steinhardt, and J. Toner, \emph{Elasticity and Dislocations
in Pentagonal and Icosahedral Quasicrystals}, Phys. Rev. Lett. \textbf{54},
1520 (1985).


\bibitem{De1987}P. De and R. A. Pelcovits, \emph{Linear elasticity
theory of pentagonal quasicrystals}, Phys. Rev. B \textbf{35},
8609 (1987).

\bibitem{Ding1993}D-h. Ding, W. Yang, C. Hu, and R. Wang, \emph{Generalized
elasticity theory of quasicrystals}, Phys. Rev. B \textbf{48}, 7003
(1993).

\bibitem{Lian2019}B. Lian, Z. Wang, and B. A. Bernevig, \emph{Twisted
Bilayer Graphene: A Phonon-Driven Superconductor}, Phys. Rev. Lett.
\textbf{122}, 257002 (2019).

\bibitem{Fernandes2020}R. M. Fernandes and J. W. F. Venderbos, \emph{Nematicity
with a twist: Rotational symmetry breaking in a moir\'e superlattice},
Sci. Adv. \textbf{6}, eaba8834 (2020).

\bibitem{Lubensky1985} T. C. Lubensky, S. Ramaswamy, and J. Toner,
\emph{Hydrodynamics of icosahedral quasicrystals}, Phys. Rev. B \textbf{32}, 7444 (1985).

\bibitem{Baggioli2000} M. Baggioli and M. Landry,
\emph{Effective field theory for quasicrystals and phasons dynamics},
SciPost Phys. \textbf{9}, 062 (2020).

\bibitem{Ochoa2021} H. Ochoa and R. M. Fernandes, unpublished.

\bibitem{Pretko2018}M. Pretko and L. Radzihovsky, \emph{Fracton-Elasticity
Duality}, Phys. Rev. Lett. \textbf{120}, 195301 (2018).

\bibitem{Pai2018}S. Pai and M. Pretko, \emph{Fractonic line excitations:
An inroad from three-dimensional elasticity theory}, Phys. Rev. B
\textbf{97}, 235102 (2018).

\bibitem{Pretko2019}M. Pretko, Z. Zhai, and L. Radzihovsky, \emph{Crystal-to-fracton
tensor gauge theory dualities}, Phys. Rev. B \textbf{100}, 134113
(2019).

\bibitem{Radzihovsky2020a}L. Radzihovsky and M. Hermele, \emph{Fractons from Vector Gauge Theory}, 
Phys. Rev. Lett. \textbf{124}, 050402 (2020).

\bibitem{Gromov2019}A. Gromov, \emph{Chiral Topological Elasticity and Fracton Order}, Phys. Rev. Lett. \textbf{122}, 076403 (2019).

\bibitem{Gromov2020b}A. Gromov and P. Surowka, \emph{On duality between Cosserat elasticity and fractons}, SciPost Phys. \textbf{9}, 076 (2020).

\bibitem{Nguyen2020}D. X. Nguyen, A. Gromov and S. Moroz, \emph{Fracton-elasticity duality of two-dimensional superfluid vortex crystals: defect interactions and quantum melting}, SciPost Phys. \textbf{9}, 076 (2020). 









\bibitem{Louie2010} O. V. Yazyev and S. G. Louie, \emph{Topological defects in graphene: Dislocations and grain boundaries}, Phys. Rev. B \textbf{81}, 195420 (2010).

\bibitem{Kirkland2012} J. H. Warner, E. R. Margine, M. Mukai, A. W. Robertson, F. Giustino, and A. I. Kirkland, \emph{Dislocation-Driven Deformations in Graphene}, Science \textbf{337}, 209 (2012).

\bibitem{Butz2013} B. Butz, C. Dolle, F. Niekiel, K. Weber, D. Waldmann, H. B. Weber, B. Meyer, and E. Spiecker, \emph{Dislocations in bilayer graphene}, Nature \textbf{505}, 533 (2014). 

\bibitem{Zeldov2020} A. Uri, S. Grover, Y. Cao, \emph{et al.}, \emph{Mapping the twist-angle disorder and Landau levels in magic-angle graphene}, Nature \textbf{581}, 47 (2020).

\bibitem{Bediako2021} N. P. Kazmierczak, M. Van Winkle, C. Ophus, K. C. Bustillo, S. Carr, H. G. Brown, J. Ciston, T. Taniguchi, K. Watanabe, and D. K. Bediako, \emph{Strain fields in twisted bilayer graphene}, Nat. Mater. \textbf{20}, 956 (2021).

\bibitem{Nelson1979}D. Nelson and B. I. Halperin, \emph{Dislocation-mediated
melting in two dimensions}, Phys. Rev. B \textbf{19}, 2457 (1979).

\bibitem{Young1979}A. P. Young, \emph{Melting and the vector Coulomb
gas in two dimensions}, Phys. Rev. B \textbf{19}, 1855 (1979).

\bibitem{Kleinert1989}H. Kleinert, \emph{Gauge fields in Condensed
Matter Physics, Stresses and Defects, Differential Geometry, Crystal
Defects}, World Scientific, Singapore, (1989).

\bibitem{Zaanen2004}J. Zaanen, Z. Nussinov, and S. I. Mukhin, \emph{Duality
in 2 + 1D quantum elasticity: superconductivity and quantum nematic
order}, Annals of Physics \textbf{310}, 181 (2004).

\bibitem{Cvetkovic2007}V. Cvetkovi\'c, Z. Nussinov, and J. Zaanen,
\emph{Topological kinematic constraints: dislocations and the glide
principle}, Philosophical Magazine \textbf{86}, 2995 (2007).

\bibitem{Kleinert2008} H. Kleinert, \emph{Multivalued Fields
In Condensed Matter, Electromagnetism, and Gravitation}, World Scientific, Singapore, (2008).

\bibitem{Bohsunng1987}J. Bohsung and H.-R. Trebin, \emph{Disclinations in quasicrystals},
Phys. Rev. Lett. \textbf{58}, 1204  (1987); Erratum Phys. Rev. Lett. \textbf{58}, 2277 (1987).

\bibitem{Kleman2003}M. Kleman, \emph{Phasons and the plastic deformation of quasicrystals}, European Physical Journal B \textbf{31}, 315 (2003).

\bibitem{Kleman2013}M. Kleman, \emph{Defects in quasicrystals, revisited I- flips, approximants, phason defects}, arXiv:1303.5563; \emph{Defects in quasicrystals, revisited II- perfect and imperfect dislocations},   arXiv:1303.5773.

\bibitem{Nandkishore_review}R. M. Nandkishore and M. Hermele, \emph{Fractons}, Annual Review of Condensed Matter Physics, \textbf{10}, 295 (2018).

\bibitem{Chamon2005}C. Chamon, \emph{Quantum Glassiness in Strongly
Correlated Clean Systems: An Example of Topological Overprotection},
Phys. Rev. Lett. \textbf{94}, 040402 (2005).

\bibitem{Haah2011}J. Haah, \emph{Local stabilizer codes in three
dimensions without string logical operators}, Phys. Rev. A \textbf{83},
042330 (2011).

\bibitem{Pretko2017}M. Pretko, \emph{Generalized electromagnetism of subdimensional
particles}, Phys. Rev. B \textbf{96}, 035119 (2017).

\bibitem{Gromov2020}A. Gromov, A. Lucas, and R. M. Nandkishore, \emph{Fracton hydrodynamics},
Phys. Rev. Research \textbf{2}, 033124 (2020).

\bibitem{Feldmeier2020}J. Feldmeier, P. Sala, G. De Tomasi, F. Pollmann,
and M. Knap, \emph{Anomalous Diffusion in Dipole- and Higher-Moment-Conserving Systems},
Phys. Rev. Lett. \textbf{125}, 245303 (2020).

\bibitem{Moudgalya2021}S. Moudgalya, A. Prem, D. A. Huse, A. Chan,
\emph{Spectral statistics in constrained many-body quantum chaotic systems}, arXiv:2009.11863.

\bibitem{Radzihovsky2020}L. Radzihovsky, \emph{Quantum Smectic Gauge Theory},
Phys. Rev. Lett. \textbf{125}, 267601 (2020). 

\bibitem{Kamenev2011} A. Kamenev, \emph{Field Theory of Non-Equilibrium Systems} (Cambridge University Press, 2011).

\bibitem{vdWaerden1928}B. L. Van der Waerden, 
\emph{Spinoranalyse}, Nachr. Ges. Wiss. G\"ottingen Math.-Phys. 100-109 (1928).



\bibitem{Peterson2021}C. W. Peterson, T. Li, W. Jiang, T. L. Hughes, and  G. Bahl, \emph{Trapped fractional charges at bulk defects in topological insulators}, Nature \textbf{589}, 376 (2021).

\bibitem{Surowka2021}P. Sur\'owka, \emph{Dual gauge theory formulation
of planar quasicrystal elasticity and fractons}, arXiv:2101.12234.

\bibitem{Else2021}D. V. Else, S.-J. Huang, A. Prem, and A. Gromov, \emph{Quantum many-body topology of quasicrystals}, arXiv:2103.13393.

\end{thebibliography}
\end{document}